# Systemic Risk in the International System

## *Criticality and Sensitivity of the International System*

Ingo Piepers
Tower Projects - Amsterdam

E-mail: ingopiepers@gmail.com

**Abstract:**

The risk of systemic war seems dependant on the level of criticality and sensitivity of the International System, and the system's conditions. The level of criticality and sensitivity is dependant on the developmental stage of the International System. Initially, following a systemic war, the increase of the level of criticality and sensitivity go hand in hand. However, at a certain stage the sensitivity of the International System for larger sized wars decreases; as a consequence of a network effect, we argue. This network effect results in increased local stability of the system. During this phase the criticality of the International System steadily increases, resulting in a release deficit. This release deficit facilitates a necessary build up of energy to push the International System, by means of systemic war, into a new stability domain. Systemic war is functional in the periodic rebalancing of an anarchistic international system.

## 1. Introduction

In this paper we explore a methodology to assess the potential for war - including systemic war - in the International System. According to our research it is possible to differentiate between (1) 'normal' wars - wars during the life span of international systems, and (2) periodic systemic ('general') wars. Systemic wars have some specific characteristics: e.g. their size, their high intensity, and their specific stake: the 'new' governance structure of the International System. Systemic risk concerns the probability of local and regional war(s) becoming systemic.

In order to assess the probability of normal and systemic wars, the approach discussed in this paper focuses on - what we call - the criticality and the sensitivity of the International System. Our research shows that a relationship exists between, on the one hand, the criticality and sensitivity of the International System, and on the other hand, the developmental stage ('age') of the International System; that is, its development stage in the typical life cycle of the system.

We make use of various concepts related to complexity science - especially network theories - and of results of our earlier exploratory research.

We will start with the definitions of some key concepts and with an explanation of the exploratory methodology we use to assess the level of criticality and sensitivity of the International System. Next we will discuss the vulnerable clusters we have identified in the current International System. Than we will try to qualify and quantify the current war potential - the systemic risk - of the International System. Finally we will do some suggestions for further research.

## 2. <u>Methodology</u>

### 2.1 Introduction

In this paragraph we discuss the methodology to assess the potential for war - including systemic war - in the International System. This perspective is based on our earlier research findings, and on research by Watts, concerning the 'cascade properties' in a network with varying degrees of connectivity, and with threshold influencing the decisions of agents.
Relevant assumptions based on earlier exploratory research are (see for a more extensive explanation Appendix 1):

- The requirement for social systems, including states and the International System itself, to fulfill an interdependent set of basic functions, in order to function properly – with a certain efficiency – and to survive.
- The self-organized criticality (SOC) characteristics, and resulting punctuated equilibrium (PE) dynamics of the International System.
- The - at least to a certain degree – chaotic war dynamics of the International System.
- The functionality of SOC- and PE- (war) dynamics in the increase in stability of the International System, contributing to a process of social consolidation.

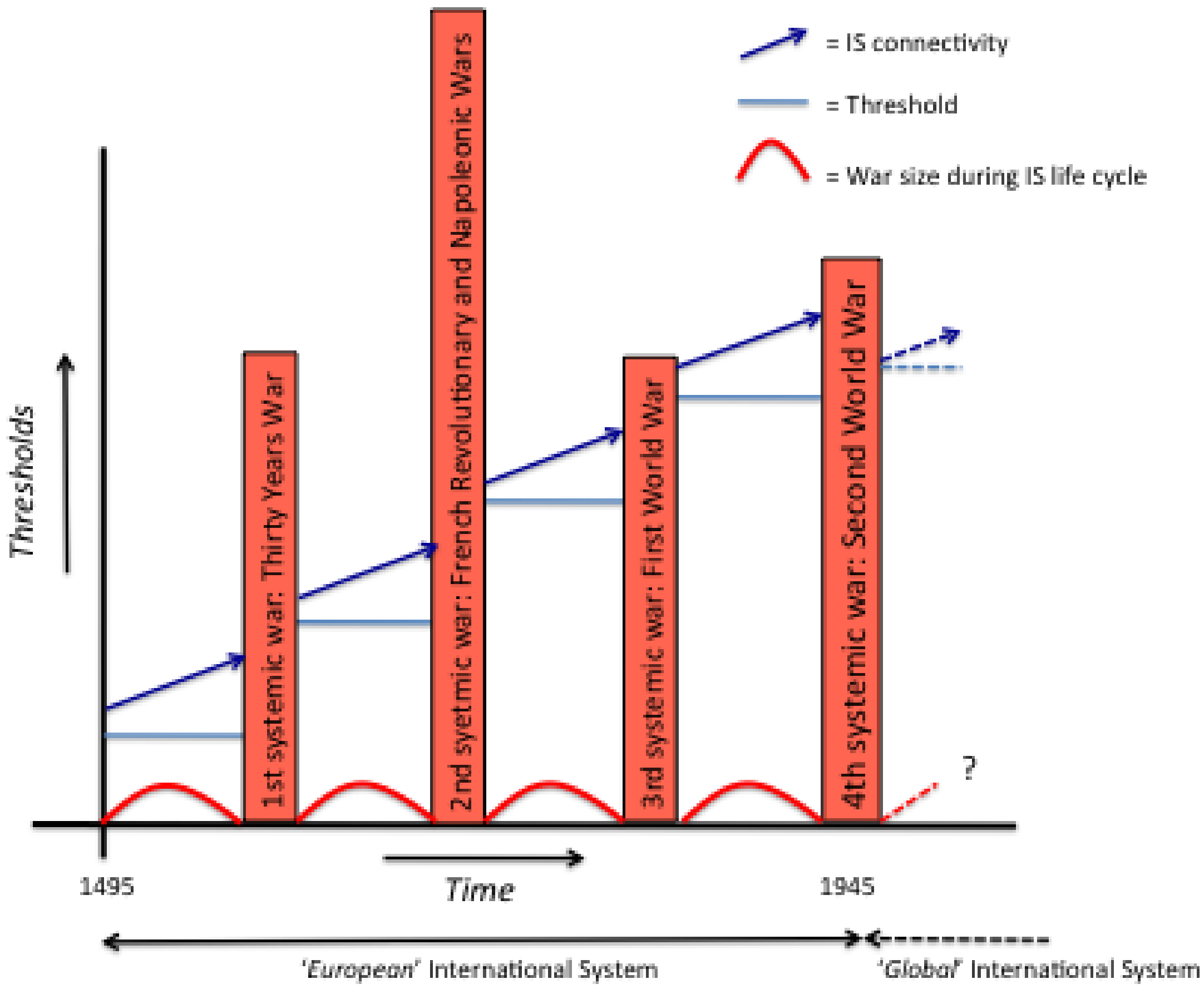

*Figure 1. This figure shows the four systemic wars that took place during the period 1495 - 1945, and the development of various characteristics of the International System. See Appendix 1, for further explanation.*

As we will explain in this paragraph, three properties of the International System and their development seem to influence the war dynamics of this system:

- *The structure and the size - the cascade condition - of the network of so called 'vulnerable clusters', and its development over time*. The configuration of vulnerable clusters determines the level of *criticality* of an international system.
- *The sensitivity of the International System for war.* In our approach sensitivity is related to the size of wars: Increased war size over time implies greater sensitivity. A relationship seems to exist between the sensitivity of an international system (war size), and its development stage.
- *Various conditions and developments in the system's environment.* These conditions have in common that they affect the ability of states and the International System to fulfill their basic functions.

Criticality of the International System concerns the dynamics of the network of vulnerable clusters; sensitivity concerns the dynamics on the network of vulnerable clusters: wars.
Our research suggest that a more critically organized international system, not necessarily is more sensitive for war: Criticality and sensitivity of the International System do not necessarily go hand in hand, as we will explain. At a certain stage in the life cycle of an international system, a network effect starts to hinder the war dynamics of the system, decreasing its sensitivity.
From this perspective three questions need answering to assess the current probability of (systemic) war:

- What is the current level of criticality of the International System, in other words, what is the structure and size of vulnerable clusters in the International System?
- What is the current level of sensitivity of the International System? In order to answer this - and the preceding question - question we will make use of findings of earlier research, suggesting that consecutive international systems - relatively stable periods in between two systemic wars - have a typical life cycle.
- What is the influence of specific conditions on the level of criticality and sensitivity of the International System?

## 2.2 Criticality of the International System

For the definition and operationalization of the concept of criticality of the International System, we make use of the theoretical research of Watts[1] (See appendix 2).
The theoretical research of Watts suggests that two (network) characteristics of his model influence the size and number of cascades: (1) the connectivity of the system, and (2) thresholds[2] that define when a node switches its 'position'. In this context states are considered nodes, and a decision by a state to use violence against another state, or to actively participate in a war, as a 'change in position'.
We suggest – based on our research – and the characteristics in the war dynamics of the International System in the time frame 1495 – 1945, that more or less identical properties – the connectivity of the International System and thresholds in the International System – play a significant role in the dynamics of this real-life system as well.

[1] Watts explains the phenomenon of rare cascades that are triggered by small initial shocks, "in terms of a sparse, random network of interacting agents whose decisions are determined by the actions of their neighbors according to a simple threshold rule." (Watts, 2002, 5766)

[2] According to Watts, "Because many decisions are inherently costly, requiring commitment of time or resources, the relevant decision function frequently exhibits a strong threshold nature: agents (decision makers of states in this context, IP) display inertia in switching states, but once their personal threshold has been reached, the action of even a single neighbor can tip them one state (no use of violence against another state, IP) to another" (use of violence is seen as practical and/or justified, IP) (Watts, 2002, 5767).

Connectivity of the International System concerns the number of connections between states in the International System. We assume that the connectivity of the resulting network increases during its life cycle.
Thresholds in the International System consist of rules and institutions that make up its governance structure, and concern the use of violence (war) against other states. Interests of states and institutions probably contribute to a threshold effect as well.
We argue that during the period 1495 – 1945, the thresholds of consecutive international systems have increased, e.g. restricting the legitimacy of the use of violence (Piepers, 2007).

### 2.3 Sensitivity of the International System

Watts model suggests, that at a certain stage, the connectivity of a network can hinder cascades in a system. The reason for this typical dynamic is the increased local stability of the network, hindering cascades to progress.
As discussed, our research shows a more or less similar dynamic: initially - following a systemic war - the size of wars increases, than a peak is reached, and the size of wars decreases until the next - abrupt, but not unexpected - systemic war. We assume that a more or less similar mechanism - a network effect - is the cause of this typical war dynamic over time.

This means - according to our definition - that the sensitivity of an international system initially increases, reaches a peak, decreases and than suddenly dramatically increases, allowing for the next systemic war.
Below figure shows this typical dynamic, in combination with the steady increase in the level of criticality of the International System.

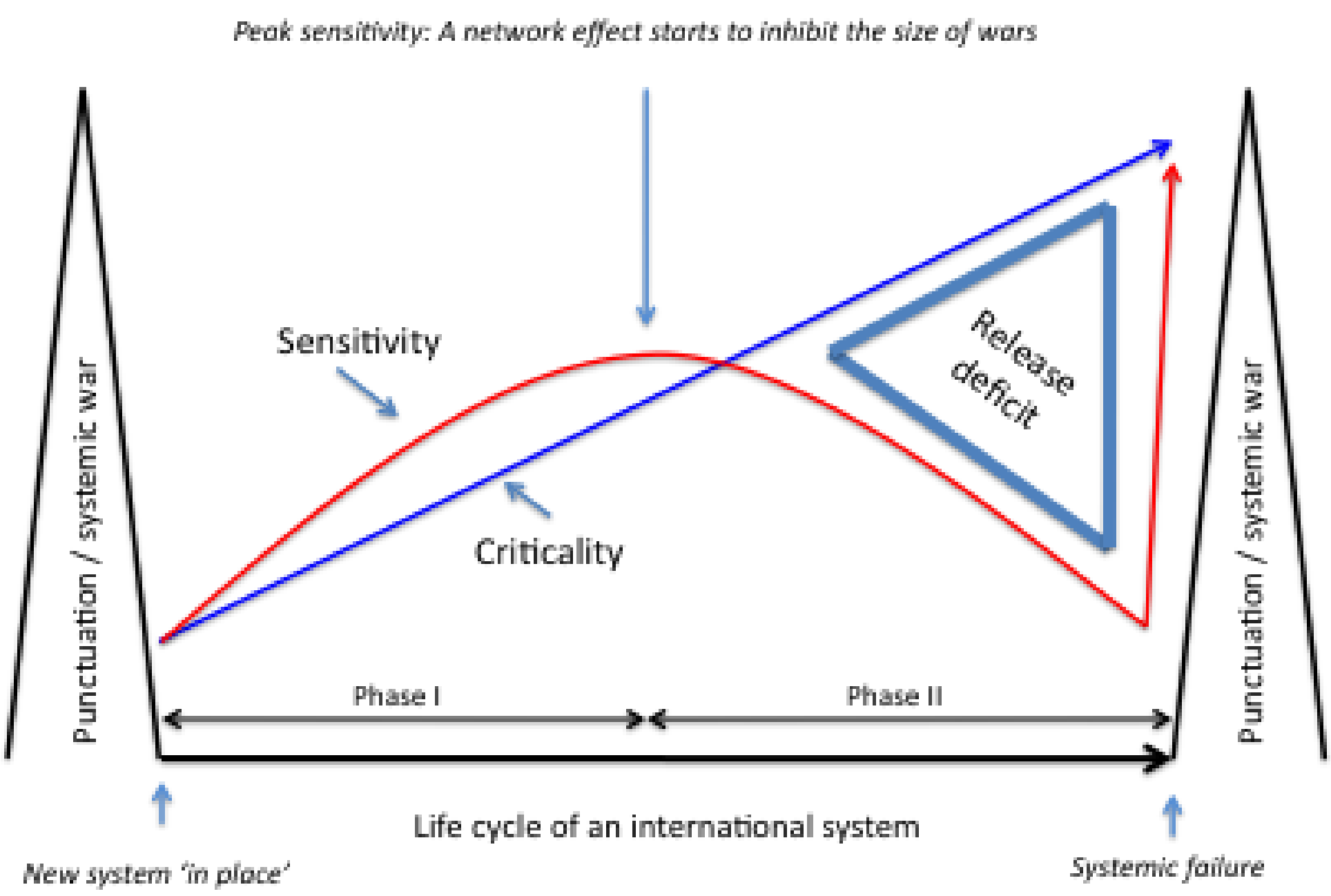

*Figure 2. This figure shows the development of the level of criticality and sensitivity during the life cycle of an international system.*

### 2.4 Conditions of the International System

Various conditions and their development influence the (war) dynamics of the International System. We consider the set of basic functions - and the requirement that these functions need fulfillment and balancing - the mechanism, that to a high degree determines the effects of these conditions and developments.

Examples that probably have these effects - through the workings of the basic functions - on the war dynamics are: population growth (Piepers, 2007), climate change and its effects, and resource scarcity.

### 2.5 Synthesis: A framework to assess the war potential of the International System

Based on this typical behavior, we differentiate between two phases during the life cycle of an international system[3].

During phase I, the level of criticality and sensitivity of the International System increase more or less synchronized: tension and frustration are dealt with through release events (wars), without serious delay. However, at a certain stage, as the International System becomes more connected and issues become more entangled, states become in their dealings with other states more risk averse and restrained, it seems. 'Rational' cost/benefit calculations now become even more problematic due (partly) by a lack of transparency, and the requirement to balance even more interests; decision makers become more prudent. We suggest that these developments probably make up the network effect.

More or less simultaneously, (global) governance becomes less effective: institutions and various arrangements do not function as they are supposed to; their legitimacy is 'progressively' questioned. Great Power dynamics – their relative rise and decline – are at the heart of these processes; fueling a sense of insecurity and anxiety. States start to realize that the system becomes obsolete, and sooner or later requires rebalancing. How these imbalances should be tackled is not clear, and contributes to the criticality of the International System.

During phase II, the level of criticality of the International System further increases. But now – as explained – the war dynamics are progressively inhibited, resulting in the build up of tension and frustration. We call this build up the '*release deficit'* of the International System.

From a system perspective the build up of the 'release deficit' in fact is the build up of enough 'energy' to push the International System into another stability domain, by means of systemic war.

It is interesting to note, that the regular increase in the stability of consecutive international systems during the period 1495 - 1945, goes hand in hand with an increase in the intensity of (consecutive) systemic wars: A more stable (international) system requires more energy to rebalance and to be 'pushed' into another stability domain.

According to this concept it is possible to assess systemic risk by 'measuring' the system's level of criticality and its sensitivity, and to identify system conditions, that influence both properties.

## 3. <u>Results</u>

### 3.1 Introduction

In this paragraph we measure the potential for war in the International System, based on the concept of criticality, sensitivity and conditions, discussed in the preceding paragraph.

---

[3] This life cycle does not include the two systemic wars, respectively prior to this typical life cycle, and abruptly (but not really unexpected, as we argue) ending it.

## 3.2 Measuring criticality

In this research we define two types - levels - of vulnerable clusters. The difference of both clusters is determined by their 'vulnerability condition' (Watts, 2002, 5767).

- ***First order vulnerable clusters (1st order VC's)***. A 1st order VC consists of two or more states between which war is - or can become - a probable / a likely course of action. Criteria to qualify as 1st order VC are: explicit threats between states, arms build-up(s), alliance dynamics, and participation in proxy wars. A 1st order vulnerable cluster is unstable in a one-step sense.

- ***Second order vulnerable clusters (2nd order VC's)***. A 2nd order VC - or underlying - consists of a set of states that have a 'stake' in states making up a 1st order VC, or in the 1st order cluster itself. These states can become actively involved in the war dynamics of a 1st order VC (for example as a consequence of alliance obligations). This involvement is the effect of the workings of one or more contagion mechanisms[4]. A 2nd order vulnerable cluster is unstable in a two-step sense.

| Vulnerable clusters in de current international system | | | |
|---|---|---|---|
| | 1st order VC | | 2nd order VC |
| 1 | US | Iran | EU, ME[5], Russia, China, Pakistan, India, Afghanistan |
| 2 | US | Afghanistan | EU, ME, Russia, China, Pakistan, India, Canada, Australia, Central Asian states |
| 3 | US | Iraq | EU, ME, Russia, China |
| 4 | US | Pakistan | EU, ME, Russia, China, ME, India, Afghanistan, Australia, Central Asian states, Canada |
| 5 | Israel | Iran | US, EU, ME, Russia, China |
| 6 | Israel | Palestinians | US, EU, ME, Russia, China |
| 7 | Pakistan | India | US, EU, Russia, China, Afghanistan |
| 8 | Russia | Ukraine | US, EU, Caucasus |
| 9 | China | Taiwan | US, Russia, Japan |
| 10 | North-Korea | South-Korea | US, Russia, China, Japan |

*Table 1. This table provides an overview of 1st and 2nd order vulnerable clusters, identified in the current international system.*

In below figure the 1st order vulnerable cluster and underlying 2nd order vulnerable cluster of the United States and Iran is shown. For the other vulnerable clusters, see appendix 3.

[4] States can become part of a 1st and/or 2nd order VC through various 'contagion mechanisms'. Our assumption is that these mechanisms are closely related to the fulfillment of the basic functions of states and the IS: Survival requires fulfillment. See appendix 1.
[5] The Middle East (ME) consists of: Egypt, Jordan, Israel, Saudi Arabia, Yemen, Oman, U.A.E. Qatar, Kuwait, Iraq, Iran, Syria, Lebanon, Palestinians. The EU includes Turkey. The Caucasus consists of Georgia, Azerbaijan and Armenia. The Central Asian states are: Turkmenistan, Uzbekistan, Tajikistan, Kyrgyzstan, and Kazakhstan.

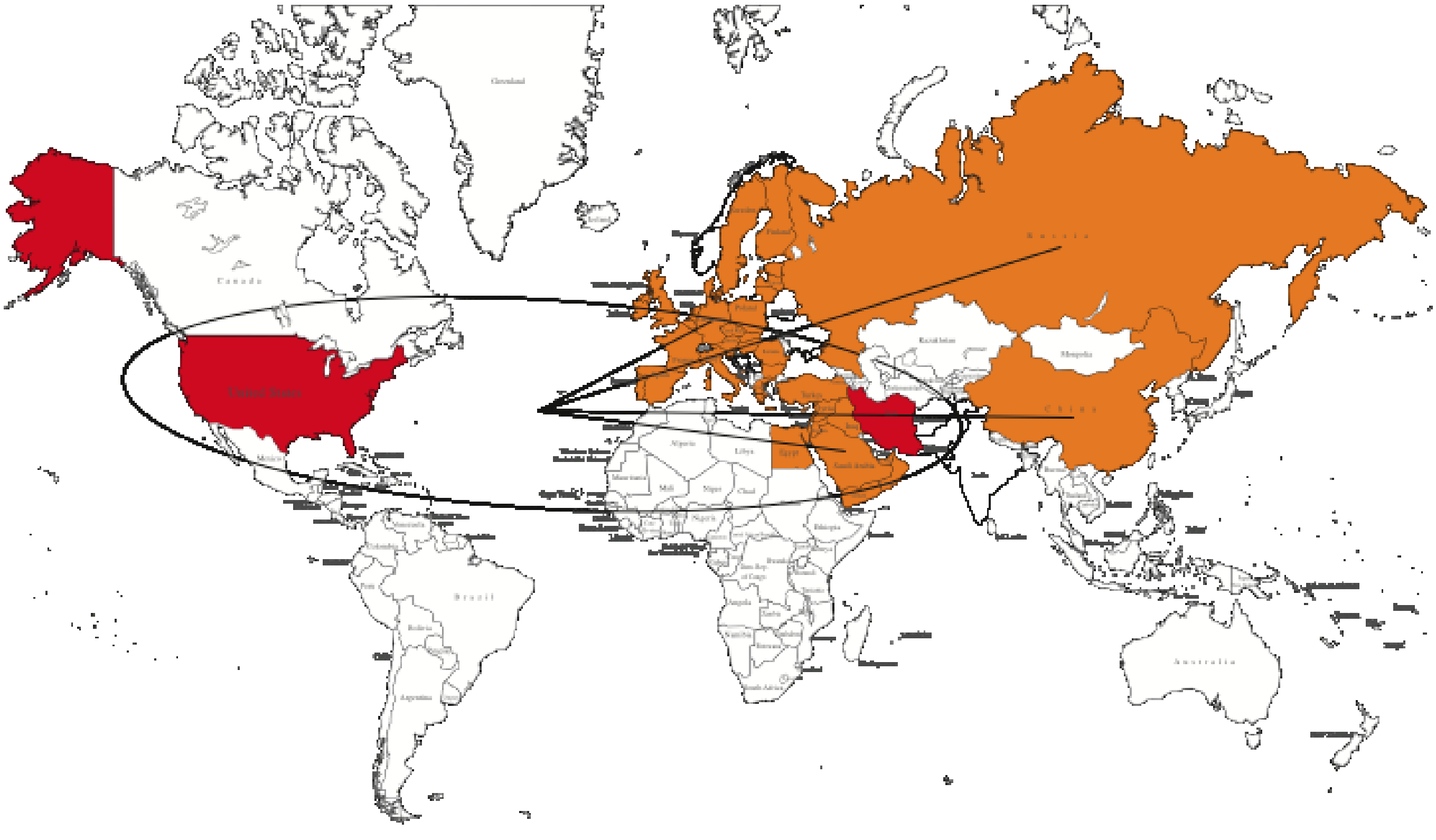

*Figure 3. The 1st (in red) and 2nd order (in orange) vulnerable clusters regarding the 'conflict' between the United States and Iran.*

This analysis makes it possible to rank states according to their degree of participation in 1st and 2nd order vulnerable clusters.

| **Connections of states to clusters, and their centrality** | | | | | | | |
|---|---|---|---|---|---|---|---|
| | **State** | **1st order VC connections** | **Centrality[6]** | **2nd order VC connections** | **Centrality** | **1st + 2nd order VC** | **Centrality** |
| 1 | US | 4 | 0,308 | 6 | 0,400 | 10 | 0,526 |
| 2 | Iran | 2 | 0,154 | 6 | 0,400 | 8 | 0,421 |
| 3 | Israel | 2 | 0,154 | 6 | 0,400 | 8 | 0,421 |
| 4 | Pakistan | 2 | 0,154 | 2 | 0,133 | 4 | 0,211 |
| 5 | Russia | 1 | 0,077 | 9 | 0,600 | 10 | 0,526 |
| 6 | China | 1 | 0,077 | 8 | 0,533 | 9 | 0,474 |
| 7 | Palestinians | 1 | 0,077 | 6 | 0,400 | 7 | 0,368 |
| 8 | Iraq | 1 | 0,077 | 6 | 0,400 | 7 | 0,368 |
| 9 | Afghanistan | 1 | 0,077 | 3 | 0,200 | 4 | 0,211 |
| 10 | India | 1 | 0,077 | 2 | 0,133 | 3 | 0,158 |
| 11 | Taiwan | 1 | 0,077 | 0 | 0,000 | 1 | 0,053 |
| 12 | Ukraine | 1 | 0,077 | 0 | 0,000 | 1 | 0,053 |
| 13 | South Korea | 1 | 0,077 | 0 | 0,000 | 1 | 0,053 |
| 14 | North Korea | 1 | 0,077 | 0 | 0,000 | 1 | 0,053 |
| 15 | EU | 0 | 0,000 | 8 | 0,533 | 8 | 0,421 |
| 16 | Canada | 0 | 0,000 | 2 | 0,133 | 2 | 0,105 |
| 17 | Japan | 0 | 0,000 | 2 | 0,133 | 2 | 0,105 |

[6] Centrality = C = deg(v) / (n-1)

| Connections of states to clusters, and their centrality (Continued) | | | | | | | |
|---|---|---|---|---|---|---|---|
| 18 | Central Asian states | 0 | 0,000 | 2 | 0,133 | 2 | 0,105 |
| 19 | Australia | 0 | 0,000 | 2 | 0,133 | 2 | 0,105 |
| 20 | Caucasus | 0 | 0,000 | 1 | 0,067 | 1 | 0,053 |
| | | 20 links | | 71 links | | 91 links | |
| | | n=14 | | n=16 | | n=20 | |

*Table 2. An overview of connections between clusters, and the resulting network centrality of states*

The US, Iran, Israel and Pakistan participate in more than one 1st order vulnerable cluster. The US is the most central node - the state with the highest centrality - of this network. As a consequence, these results suggests, the US has the highest risk of becoming involved in war. However, more factors should be taken into account to get a better understanding of the war potential of an international system.

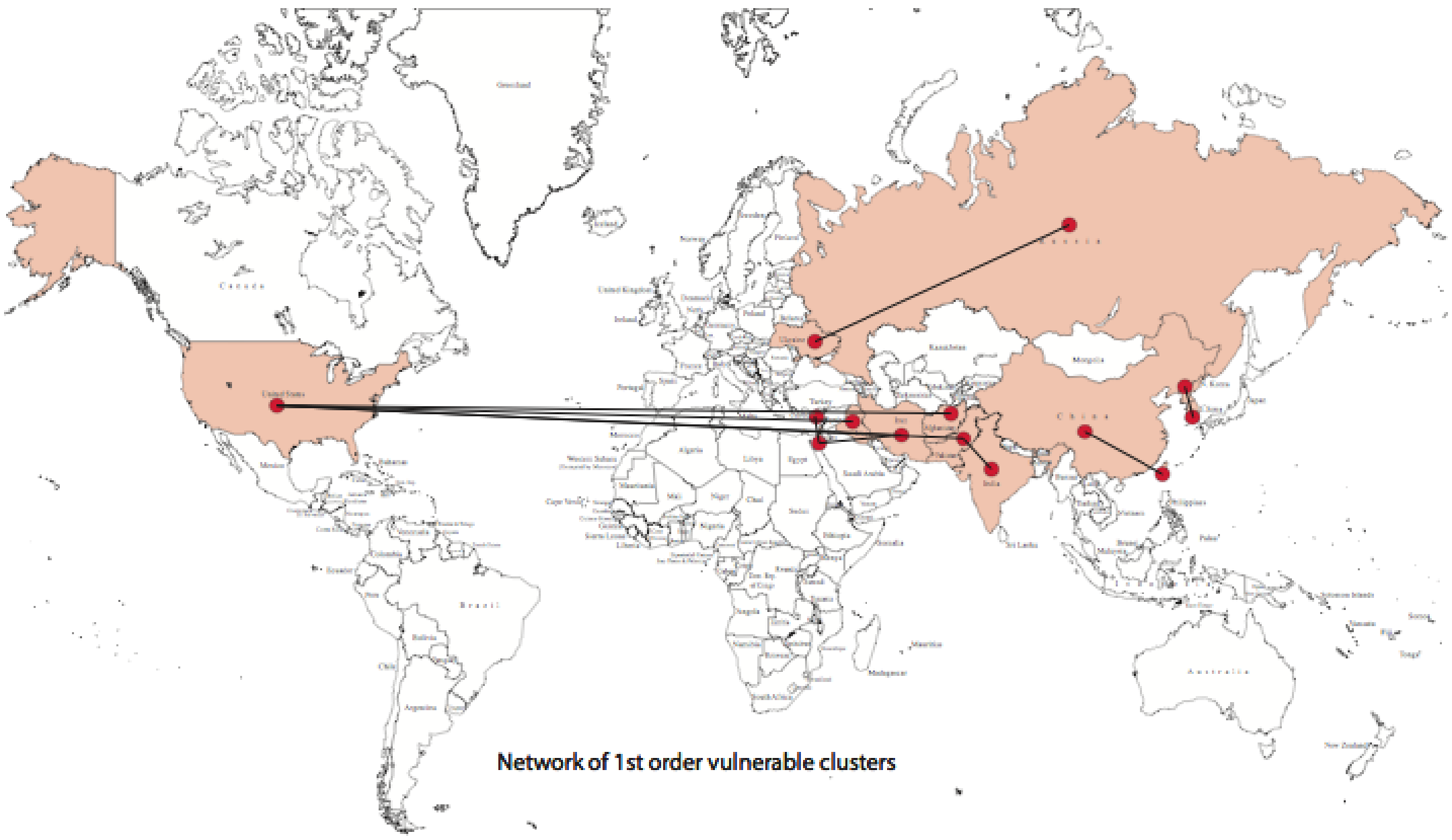

*Figure 4. Links between 1st order vulnerable clusters.*

In below schedule the causal relationships between the 1st order vulnerable clusters are further explored. This is an analysis on the level of 1st order clusters. The schedule shows with what likelihood a particular 1st order cluster can trigger another 1st order vulnerable cluster. H, M and L respectively stand for High, Medium and Low risk of contagion. This approach assigns weights to the links between 1st order vulnerable clusters.

| Contagion of 1st order vulnerable clusters | | | | | | | | | | | | |
|---|---|---|---|---|---|---|---|---|---|---|---|---|
| | 1st order VC | | 1 | 2 | 3 | 4 | 5 | 6 | 7 | 8 | 9 | 10 |
| 1 | US | Iran | | H | H | M | L | L | H | H | H | L |
| 2 | US | Afghanistan | H | | M | H | L | L | M | L | M | L |
| 3 | US | Iraq | H | M | | M | L | L | H | L | M | L |
| 4 | US | Pakistan | M | H | M | | L | L | M | L | H | L |
| 5 | China | Taiwan | L | L | L | L | | L | L | L | L | L |
| 6 | Russia | Ukraine | L | L | L | L | L | | L | L | L | L |
| 7 | Israel | Iran | H | M | H | M | L | L | | M | L | L |
| 8 | Israel | Palestinians | H | L | L | L | L | L | M | | L | L |
| 9 | Pakistan | India | H | M | M | H | L | L | L | L | | L |
| 10 | North-Korea | South-Korea | L | L | L | L | L | L | L | L | L | |

*Table 3. Degrees of contagion between vulnerable clsuters.*

This matrix enables us to quantify which clusters - when activated - have the highest risk to trigger - activate - other 1st order vulnerable clusters. The qualifications (H, M or L) give in fact a certain weight to links between two vulnerable clusters.

In below table the 'contagion risk' of 1st order vulnerable clusters is quantified. H, M and L lead to a 'score' of respectively, 3, 2 and 1 points.

| Contagion risk /escalation potential of 1st order vulnerable clusters | | | |
|---|---|---|---|
| | 1st order vulnerable cluster | | Score |
| 1 | US | Iran | 20 |
| 2 | US | Afghanistan | 16 |
| 3 | US | Iraq | 16 |
| 4 | US | Pakistan | 16 |
| 5 | China | Taiwan | 9 |
| 6 | Russia | Ukraine | 9 |
| 7 | Israel | Iran | 16 |
| 8 | Israel | Palestinians | 12 |
| 9 | Pakistan | India | 15 |
| 10 | North-Korea | South-Korea | 9 |

*Table 4. In this table the contagion risk - escalation potential of the various 1st order vulnerable clusters is quantified; 9 is the lowest possible score. Clusters with this score can be typified as 'stand alone' clusters.*

It is possible to rank states on the basis of the contagion risk / escalation potential of the 1st order vulnerable clusters they participate in (this is an analysis on the level of states). For example, both the US and Iran score 20 points because of their participation in the first 1st order vulnerable cluster, etc..

| Contagion risk / escalation potential of states | | |
|---|---|---|
| | **State** | **Score** |
| 1 | US | 68 |
| 2 | Iran | 36 |
| 3 | Pakistan | 31 |
| 4 | Israel | 28 |
| 5 | Iraq | 16 |
| 6 | Afghanistan | 16 |
| 7 | India | 15 |
| 8 | Palestinians | 12 |
| 9 | Russia | 9 |
| 10 | China | 9 |
| 11 | Taiwan | 9 |
| 12 | Ukraine | 9 |
| 13 | South Korea | 9 |
| 14 | North Korea | 9 |

*Table 5. In this table states are rank according to their 'individual' contagion risk. It is important to note that the states intentions and interest are not taken into account in this analysis.*

This analysis shows that (especially) the US, Iran, Pakistan and Israel are vulnerable states, that can 'easily' cause conflicts to escalate.
However, so far we did not include in our analysis the fact that states have different interests. They not only vary in their 'own' thresholds for the use of violence against other states, and in their willingness to comply with the thresholds established by - for example - the United Nations.
The interests that are at stake for the various states in (potential) conflicts differ fundamentally, as well.
For Iran, Israel and North Korea the survival of the state - or at least its leadership - is at stake. For the United States, the 'American' International System, its favorable arrangements, and its dominant position in the current system, are at stake. For Iran, Israel, and North Korea escalation could pay off.
Iran not only is the main challenger of the system, continuously provoking other states and questioning its legitimacy, but also the main threat to the existence of the state of Israel.
Although the United States is involved in four 1$^{st}$ order and six 2$^{nd}$ order vulnerable clusters, the United States has an interest in avoiding contagion and escalation of conflicts and vulnerable clusters: (simultaneous) activation of these clusters will further affect its dominant position in the International System.
Iran, on the other hand, threatens the United States - recognizing this particular vulnerability - with escalation, in case one of its two 1$^{st}$ order vulnerable clusters is activated. Iran has build up its potential to interfere in other 1$^{st}$ order vulnerable clusters: in Iraq and through Hezbollah and Hamas in the Near East (respectively through clusters 3 and 6).
The network of 2$^{nd}$ order vulnerable clusters has some interesting characteristics as well. Russia, China and the EU are the most central nodes in this network. However, these states are not involved in more than one 1$^{st}$ order vulnerable clusters. The EU only participates in 2$^{nd}$ order vulnerable clusters, despite its global interests; illustrative for its introvert policies, and (resulting) role as follower.
The US, Iran and Israel have a significant number of links with 2$^{nd}$ order vulnerable clusters as well. The network of 2$^{nd}$ order vulnerable clusters can be instrumental in spreading conflicts.

### 3.3 Measuring sensitivity

In paragraph 2 we have proposed a concept to 'measure' the sensitivity of the current International System. It should be stressed that this is a highly speculative endeavor, because of the following reasons:

- A (still speculative) conclusion from our previous research is that - during the time frame 1495 -1945 - four consecutive international systems can be identified, with remarkable qualitative and quantitative regularities in its (war) dynamics. However, we have argued (Piepers, 2007) that from 1945 onwards, a new - global - system has emerged. This global international system is not dominated exclusively by European Great Powers. The Second World War was in this sense a turning point. The question is: does this new - Global - International System have the same system logic as its European predecessor? By 'system logic' we refer to the typical long term dynamic in which relatively stable periods ('international systems') are more or less regularly punctuated by systemic wars, in order to rebalance the system and create new potential for growth and development? This question cannot be answered: not enough data is available about the current international system to identify the characteristics of its war dynamics.
- The chaotic war dynamics of the International System are another complication. It is not possible to make sense out of recent war dynamics, as a consequence of the chaotic attractor (after 1991).
- Finally, it should be noted that the normal - chaotic - war dynamics of the current international system were probably disturbed by the ossification of the dynamics and development of the International System during the Cold War; resulting in another complication in the assessment in recent war dynamics (see Appendix 1).

When determining the development stage - the sensitivity - of the current system, indications contradict each other. Assuming that the current international system develops according to the same logic, as did its predecessors, our conclusion is that the system is still in phase I, during which the size of wars increases. In other words, the network effect is not yet inhibiting the war dynamics of the current system. The analysis in the preceding paragraph supports this observation: a number of Great Powers (China, Russia and Brazil) will probably not participate in a escalating war, even in case - maybe we should say 'especially in case' - the first four cluster are activated (letting the US exhaust itself even more).
On the other hand, states seem to become more frustrated with the functioning of the current international system, the current arrangements of the system are losing their legitimacy: its institutions (e.g. the composition of the Security Council of the United Nations), and certain rules (the Non Proliferation Treaty, and its selective interpretation), are more and more contested by 'new' Great Powers. The question is if the current system has - despite these frustrations and resulting tension - enough 'staying power', and is not yet in phase II.

### 3.4 Indentifying conditions and their development

We have suggested that three conditions have - by means of the basic functions - an (in)direct impact on the criticality and sensitivity of the International System. It is important to stress that these causal relationships imply that these conditions have an impact on the life span of international systems as well. We have explored the relationship between the life span of international systems, and population growth; such a relationship probably exists (Piepers, 2007).
The world's population is still growing, resource scarcity increases rapidly, and climate change - through its (in)direct effects, creates tension in the international system, as well. These developments increase the risk of war.

## 4. Conclusions

Based on the framework developed in this paper, we conclude that the war potential of the current international system has not yet reached its peak, and still is increasing. In case the current international system develops according to the same life-cycle logic as we have identified in the four systems (in the time frame 1495 -1945), than at a certain stage the size of wars will decrease, and a network effect will result in a release deficit, and in the build up of energy to re-balance the system through systemic war.
However, their seems to be an alternative: As we suggested in earlier research, the International System - as do the models in Watts' experiments - can reach a level of connectivity ruling out cascades completely. This seems to have happened in Europe (Piepers, 2008).

## 5. Further research

Further research is required to validate this concept and to reach more accurate results; this is just the first step in the development of a concept to assess the war potential of the International System.

- Validation of the central assumptions of this research: the workings and effects of basic functions, SOC-characteristics of the International System, PE-dynamics, Chaotic war dynamics, the existence of a life cycle, and the functionality of wars in a process of social consolidation.
- Validation of the supposed 'regular' increase of vulnerable clusters during the life cycle of an international system, by 'measuring' their number, size and configuration over time. Historical research - e.g. the development of vulnerable clusters preceding the First World War - is useful as well.
- Analysis of the effects of system conditions on the criticality and sensitivity of the International System.
- Development of more accurate criteria to identify 1$^{st}$ and 2$^{nd}$ order vulnerable clusters in the International System.
- Assessment of a probable relationship between the alliance dynamics in an international system, and the development of (the level of) criticality and sensitivity.
- Validation of the existence and workings of a network effect in the International System.
- Further research into the regular shortening of the life-span of consecutive international systems (Piepers, 2007).

## APPENDIX 1: A consistent framework

### 1. Introduction.

In this paragraph we discuss definitions and assumptions that are integral to this research. The following concepts will be discussed and defined:

- The International System and Great Powers;
- Basic Functions
- Self-organized criticality and punctuated equilibrium Dynamics
- Chaotic war dynamics
- Stability, resilience and social consolidation

### 2. The International System and Great Powers.

The assumptions in this research, are based on our research focused on the Great Power system, a dominant subsystem of the International System (Piepers, 2007).
Holsti defines an *international system* as "any collection of independent political entities - tribes, city-states, nations, or empires - that interact with considerable frequency and according to regularized processes. The analyst is concerned with describing the typical characteristic behaviour of these political units toward one another and explaining major changes in these patterns of interaction" (Holsti, 1995, pp. 23).
The International System is an anarchistic system and lacks top-down control. The limited control that is 'available' is often counterbalanced by bottom-up forces, for example as a consequence of the inter(actions) of states, and of random events (Solé, 2006, pp. 13).
Holsti argues that "for international relationships, anarchy means that ultimately states can rely only upon themselves for their security and other purposes". An outcome of the anarchic characteristics of the International System is a "process of action and reaction", states "accumulating arms for its insurance", that other states will interpret "as potential threats to their own security. This process of action and reaction is called the *security dilemma:* the means by which one state provides for its security creates insecurity for others" (Holsti, 1995, pp. 5).
International systems can be understood and analyzed at different 'nested' scales, as is for example the case with ecosystems as well (Solé, 2006, pp. 4). The following scales - levels of analysis - can be identified: (1) single states, (2) interactions between states, (3) the level of vulnerable clusters, as defined in this paper, (4) the level of the International System, and (5) the spatial context. The balance of power and international law are "elements of order" at the level of the International System (Holsti, 1995, pp. 7).
According to Levy, the *Great Power System* constitutes "a dominant subsystem in the International System, playing a major role in the transformation of the International System and the structuring of international order" (Levy, 1983, pp. 10). Levy argues that "the more powerful states - the Great Powers - determine the structure, major processes, and general evolution of the system" (Levy, 1983, pp. 8), and "the general level of interactions among the Great Powers tends to be higher than for other states, whose interests are narrower and who interact primarily in more restricted regional settings" (Levy, 1983, pp. 9). Great Powers can be identified accurately with the help of various criteria (Levy, 1983, pp. 16-19).
The year 1495 is generally considered the starting 'point' of the Great Power System (Levy, 1983). From around 1495 the dynamics of the International System - for example the distribution of power in the Great Power system - started to modify the expectations and (inter)actions of states constituting this system. From a network perspective it can be argued, that from that moment onwards, the connectivity of the International System allowed for spanning clusters - consisting of vulnerable clusters in the context of this research - to form.
In the period 1495-1975 Levy identified 119 Great Power wars; 114 out of these 119 Great Power wars took place before 1945. We differentiate between *Great Power wars*, wars with at least one Great Power participating - and *wars between Great Powers*, wars with at least two Great Powers participating. The second category constitutes seventy wars in the period 1495-1945 (Levy, 1983, pp. 70-73)[7].

### 3. Basic Functions.

We assume that social systems - states and international systems in this particular context - must fulfil four *basic functions* to ensure their survival in competitive conditions. Social systems must somehow provide for: (1) energy, and other necessities of life and wealth, (2) security against external and internal threats, (3) identity and self-development, and (4) consistency and direction (Piepers, 2006a). This concept is based on Boulding (Boulding, 1978).
Each basic function has a corresponding aspect system. Each aspect system not only has its typical structure, but goes accompanied with its typical 'rules' as well. These rules are closely related to the characteristics and functioning of the basic functions. Basic functions and aspect systems interact.
We assume that a reciprocal relationship exists between the functioning of the International System, and the ability of states to fulfil their basic functions: An international system can vary in the degree it is instrumental in the fulfilment of the basic functions of states.
Between (the variables/components) of basic functions and (the variables) of aspect systems, a minimal level of consistency is required to avoid dysfunctional inconsistencies and to facilitate growth and development.
As explained: basic functions - within a particular social system (for example a state) and in relation to the basic functions of other systems (for example: other states, supranational organizations, the international system) - require a certain consistency. "Unavoidable' growth and development of (other) social systems, result in inconsistencies, contradictions and imbalances.
At a certain stage these inconsistencies and their effects start to hinder the proper functioning of the social system itself (e.g. a state), and the International System. Especially uneven growth and development can result in inconsistencies. These inconsistencies can be local (in a geographical sense) and confined to a specific aspect system, or can be – or become – regional or global (in their effects).
Typical examples of inconsistencies are: financial imbalances resulting in crisis, divergent views concerning religion religious strife, and incongruent views about governance resulting in legitimacy issues. Inconsistencies and imbalances can be dealt with in various ways: through cooperation, competition, and sometimes conflict (war).
The 'preferred' modus to tackle inconsistencies depends on various factors: the conditions of the system, the stage of development of the International System, the availability of effective 'instruments' – e.g. policies, communication and rules – and the interests of states that are at stake.
As explained in paragraph 2 of this appendix, an anarchistic International System, 'creates' a security dilemma. The security dilemma has the effect of a driving force, contributing to the inconsistencies in the International System, and to the intensification of the rivalry between Great Powers. The security dilemma feeds on itself.
At a certain stage of the life cycle of an international system, its governance (structures and arrangements) become inconsistent with the actual power structure: 'New' - upcoming - Great Powers will (progressively) challenge the legitimacy of the current International System and its arrangements; they consider these arrangements as restrictive, slowing down their development and rise.
This dynamic goes hand in hand with the (unavoidable) erosion of power of 'older' Great Powers, that have played a major role in the establishment of the current international system, and its favourable arrangements for this particular category of states.
To correct for these 'unavoidable' inconsistencies in an anarchistic international system, spontaneous corrective adjustments of the organization of the system become unavoidable. States often have developed reasonably adequate mechanisms to deliberately realign their 'own' (internal) basic functions and corresponding aspect systems; especially democracies are effective in corrective adjustments. During the time frame 1495 - 1945, systemic war was the only effective
mechanism to 'enforce' change at the level of the system itself, and to provide for new potential for growth and development.

[7] In the period 1495-1945 eight Great Power wars took place outside Europe, of which two concern wars *between* Great Powers (the 'War of the American Revolution' (1778-1784) and the 'Russo-Japanese War' (1939)). The 'War of the American Revolution', was a war between European Great Powers on American soil, for this reason this war is included in the research discussed in this article. On the basis of a sensitivity analysis I have determined that this 'inclusion' has no effect on the patterns identified in this research.

| **Inconsistencies affecting the functioning of basic functions** | |
|---|---|
| **Basic Function** | **Inconsistency** |
| Energy, and other necessities of life and wealth | Inadequate availability of (critical) resources |
| Security against external and internal threats | The degree of actual and perceived threats to one's own security |
| Identity and self-development | Inconsistencies of values (ideological, religious, etc.), resulting in religious and cultural strife |
| Consistency and direction | The degree of alignment between actual power of states and their influence in the International System |

*Table 1.1. Examples of inconsistencies, hindering the functioning of basic functions*

## 4. Self-organized criticality and punctuated equilibrium dynamics.

According to our research, during the time frame 1495-1945, the International System shows self-organized critical (SOC) characteristics, resulting in a punctuated equilibrium (PE) dynamic (Piepers, 2007). Punctuations qualify as systemic wars, and are the result of *systemic failure* of the International System.
A SOC-system is a dynamic system in which a *driving force* - more or less constantly - 'pushes' the system towards a critical condition. *Thresholds* in these systems enable the accumulation of tension and frustration; this accumulated tension and frustration is released regularly, through - so called - release events. The size of these release events and their number show a typical statistical distribution: A power-law.

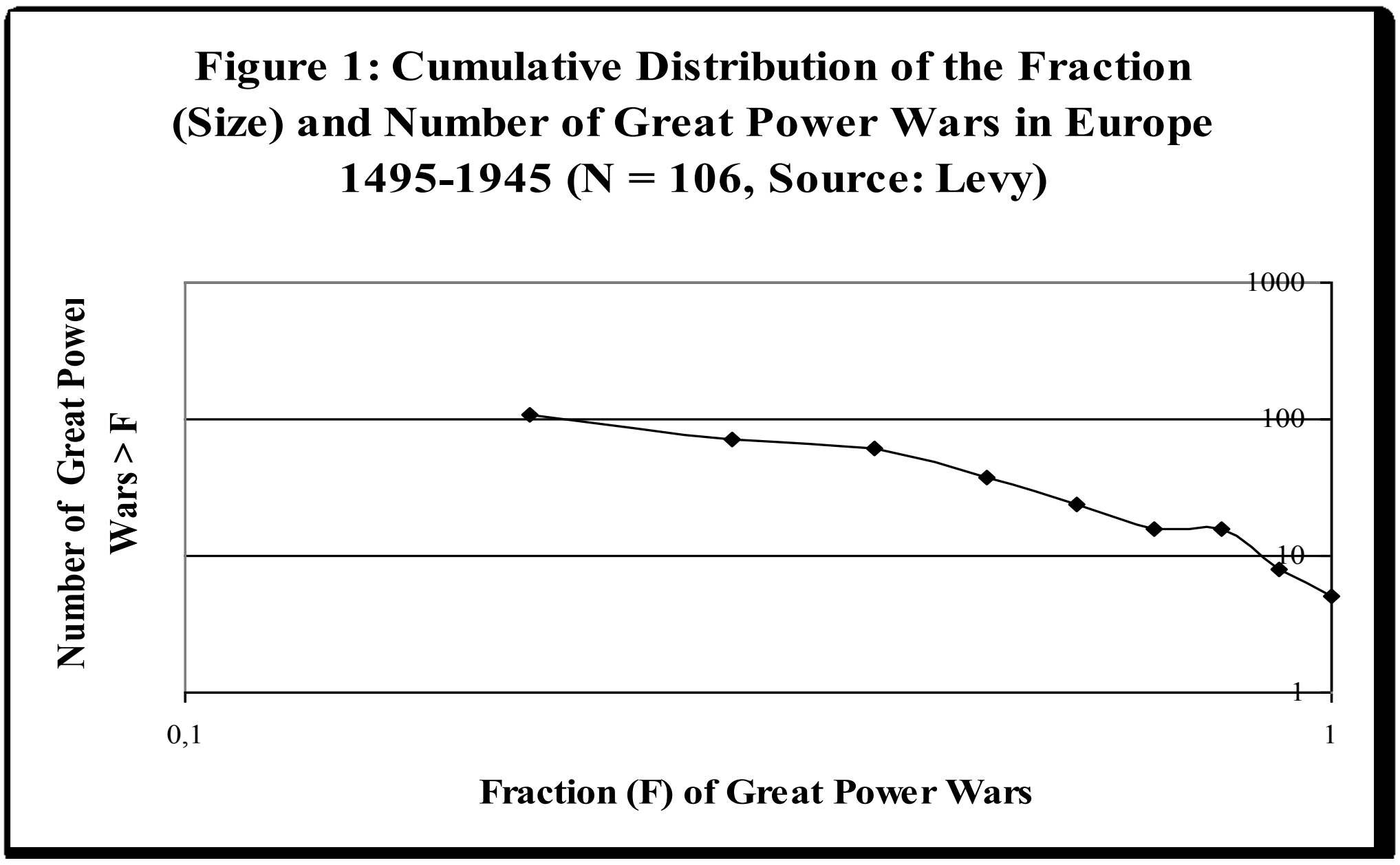

*Figure 1.1. The size of Great Power wars and their number scale as a power law (Source: Piepers, 2007)*

The regular release of tension and frustration allows for a new accumulation process. As a result, SOC produces an oscillating dynamic - a punctuated equilibrium dynamic - around the critical point of the system. These four SOC-requirements seem to be applicable for the International System[8]. During the time frame 1495 - 1945, four punctuations - systemic wars - can be identified.

[8] This means, that for SOC to be applicable, four requirements must be met. First, a critical point has to be the attractor of such a system (Bak et al. 1988, Sornette 2003, pp. 395-439, Newman 2005, pp. 12). Typically, for the dynamics of a system in a critical condition is that the size and number of release events can be shown with a power-law distribution (Newman, 2005).
Second, the system requires a driving force that more or less constantly pushes the system towards this critical point. Third, the system needs to be a threshold system. Thresholds enable the build-up of tension and frustration, and result in a necessary separation of timescales. A separation of timescales means that the accumulation of tension and frustration in the International System operates at a much slower time scale, then the release of this tension and frustration through Great Power wars (Sornette 2003, pp. 402-404). Fourth, the system needs triggers that perturb the system more or less regularly, as a result of which release events erupt, resulting - as explained - in the release of the tension and frustration that has been accumulated in the system. The International System meets these requirements.

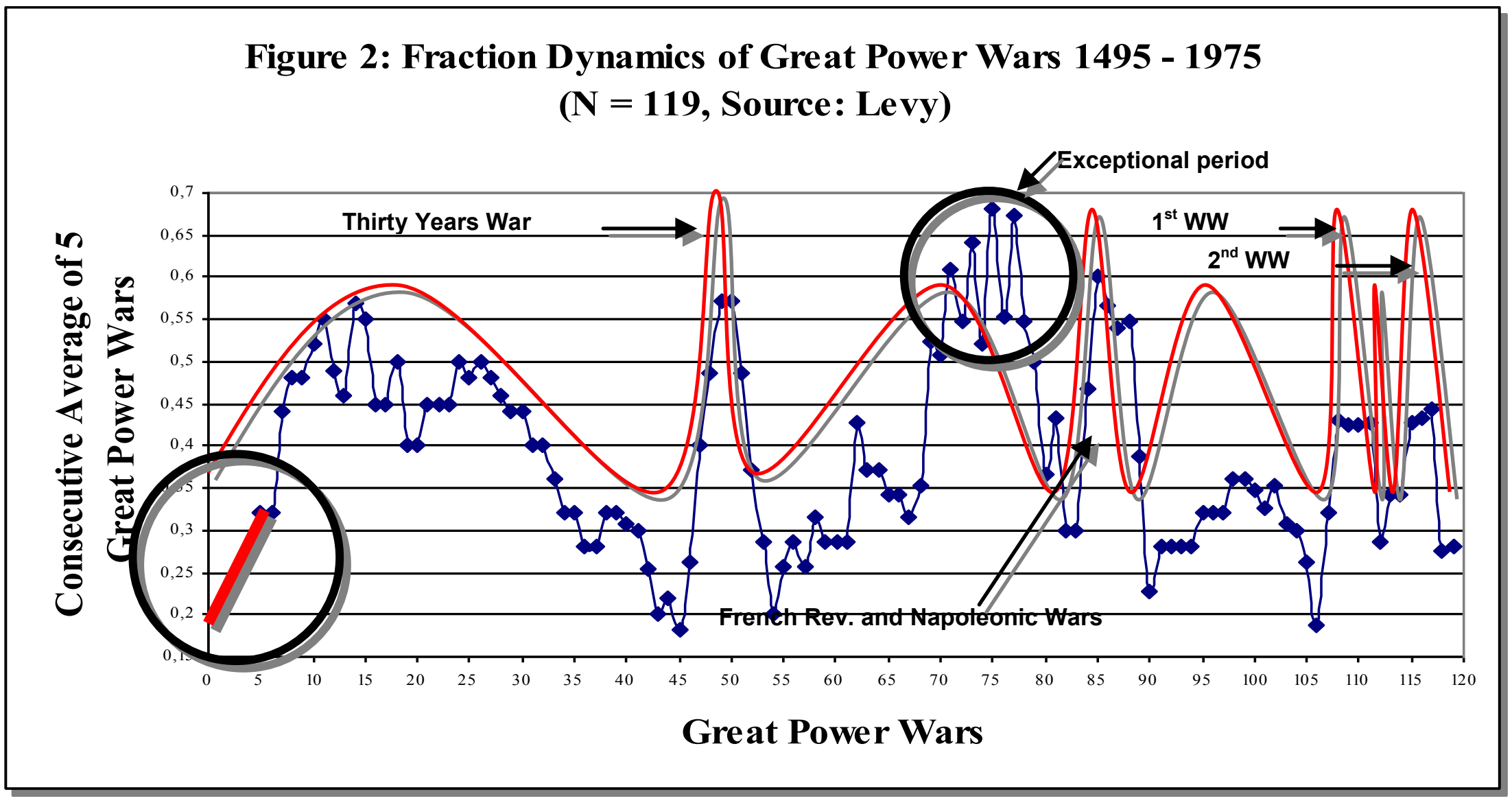

*Figure 1.2. Above figure shows four systemic wars (punctuations). However, this particular long term dynamic was disturbed during - what we call an 'exceptional period' - from 1657 - 1763. In paragraph 5 we propose a hypothesis for this distorted dynamic.*

Systemic wars - often called 'general wars' - are the result of systemic failure. In case of *systemic failure* of the International System the rules and institutions of the international system fail to stabilize the system, and fail to prevent large-scale war; the International System, in fact, collapses. In a systemic war 'new' governance of the 'new' international system is at stake.

We argue that systemic wars are fundamentally different compared to 'normal' wars, taking place during the 'regular' life cycle of international systems. Systemic wars typically have the size of the International System (including all Great Powers), have an exceptionally high intensity, and - that is the most fundamental difference - systemic wars always result in the reorganization of the International System.

Great Powers that decide systemic wars in their favor, define the new rules of the system, and typically ensure that the International System will (especially) support their own interests.

Our research shows that during the time frame 1495 - 1945, the International System was four times punctuated by systemic wars.

| **Systemic Wars** | | |
|---|---|---|
| 1 | 1618 - 1648 | Thirty Years War |
| 2 | 1792 - 1815 | French Revolutionary and Napoleonic Wars |
| 3 | 1914 - 1918 | First World War |
| 4 | 1939 - 1945 | Second World War |

*Table 1.2. Systemic wars.*

Based on this framework, it is possible to identify five consecutive international systems.

| **International Systems** | |
|---|---|
| 1 | 1495 - 1618 |
| 2 | 1648 - 1792 |
| 3 | 1815 - 1914 |
| 4 | 1918 - 1939 |
| 5 | 1945 - |

*Table 1.3. Consecutive international systems.*

Our research shows that these international systems have a typical life cycle (Piepers, 2007). Immediately after a systemic war, the size of Great Power wars always seems to have a minimal, almost zero, value[9]. Subsequently, the war fractions increase until they approach a peak. Beyond this peak the size of consecutive wars declines, and again approaches a minimal - almost zero - value. In this paper we argue that this peak can probably be attributed to a network effect.
The punctuations and life cycles, with a peak in the size of wars, makes it possible to group wars during the life cycle of an international system, in two 'war clusters'.
The first cluster consists of the wars in between the last systemic war, and the war with the largest size (the 'peak', this war is included in this cluster); in fact the wars during phase I (see paragraph 2.3 of this paper). The second cluster consists of the remaining wars during the life span of this international system (which ends abruptly with the next systemic war, Phase II wars). Wars that are part of a systemic war are not included in one of the clusters.
Because of the short life span of the fourth international system, it is not possible to identify a peak war, and war clusters.

| **Peaks/'turning points' in international systems** | | |
|---|---|---|
| **International system** | **Period** | **Peak/turning point** |
| 1 | 1495 - 1618 | 1514 |
| 2 | 1648 - 1792 | 1774 |
| 3 | 1815 - 1914 | 1856 |

*Table 1.4. Dating of peaks ' half way' the life cycle of the first international systems*

These clusters have their own 'typical' characteristics, and seem to develop according to a certain logic.

[9] The size of war is in this research is defined as the number of Great Powers participating in that particular war divided by the number of Great Powers that at that moment of time exist. Size - defined as 'fraction' - is a relative measure.

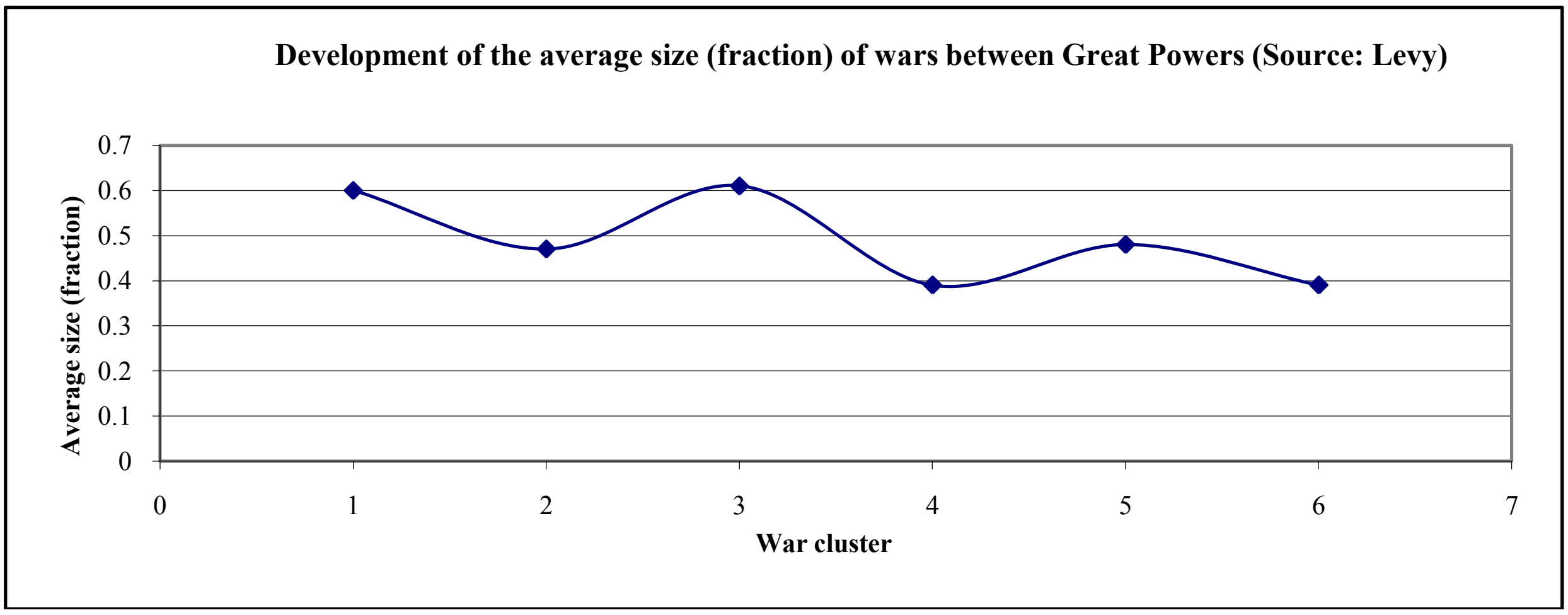

*Figure 1.3. The regular development of the average size of wars between Great Powers, based on a life cycle approach.*

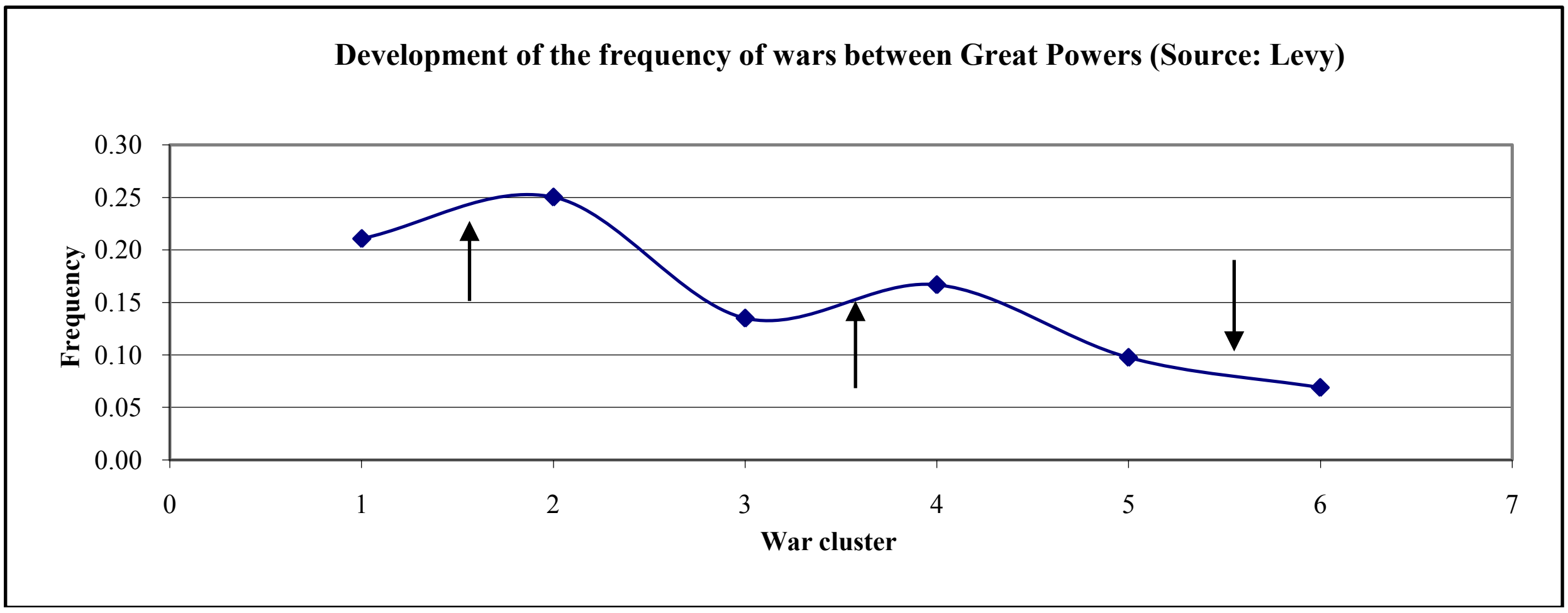

*Figure 1.4. The regular development of the frequency of wars between Great Powers, again based on a life cycle approach*

This analysis (based on Levy's dataset) gives more insight in the development of the criticality and the sensitivity of an international system during its life cycle, as well.
This analysis shows that the average size of wars always decreases during phase II of each international system; implying a decrease in sensitivity of the system, as defined in this paper.
The development of the war frequencies of war clusters (the number of wars belonging to a particular phase divided by its life span) shows an opposite development, except for the third international system.

## 5. Chaotic war dynamics

Through the construction of phase states of the war dynamics of the International System, it is possible to get a better understanding of these dynamics on the shorter term. We use two variables in these phase states: The fraction and the intensity of consecutive Great Power wars.

It seems that 'normally' the development of these two variables can be typified - and visualized - with more or less circular trajectories. A closer look reveals that sometimes these trajectories have a clockwise direction, and at other times a counter-clockwise direction (Piepers, 2007).
This analysis shows that these 'typical' trajectories were distorted during the period 1657-1763. During this specific timeframe the phase state shows a 'zigzag' pattern.
Figure 1.5 is an illustration of the phase state during the lifespan of the first international system (1495-1618). Seven circular trajectories can be identified: four with a counter-clockwise, and three with a clockwise direction[10]. During the period around 1550 and during the years from 1610 until the first punctuation (1618), these trajectories were somewhat distorted.

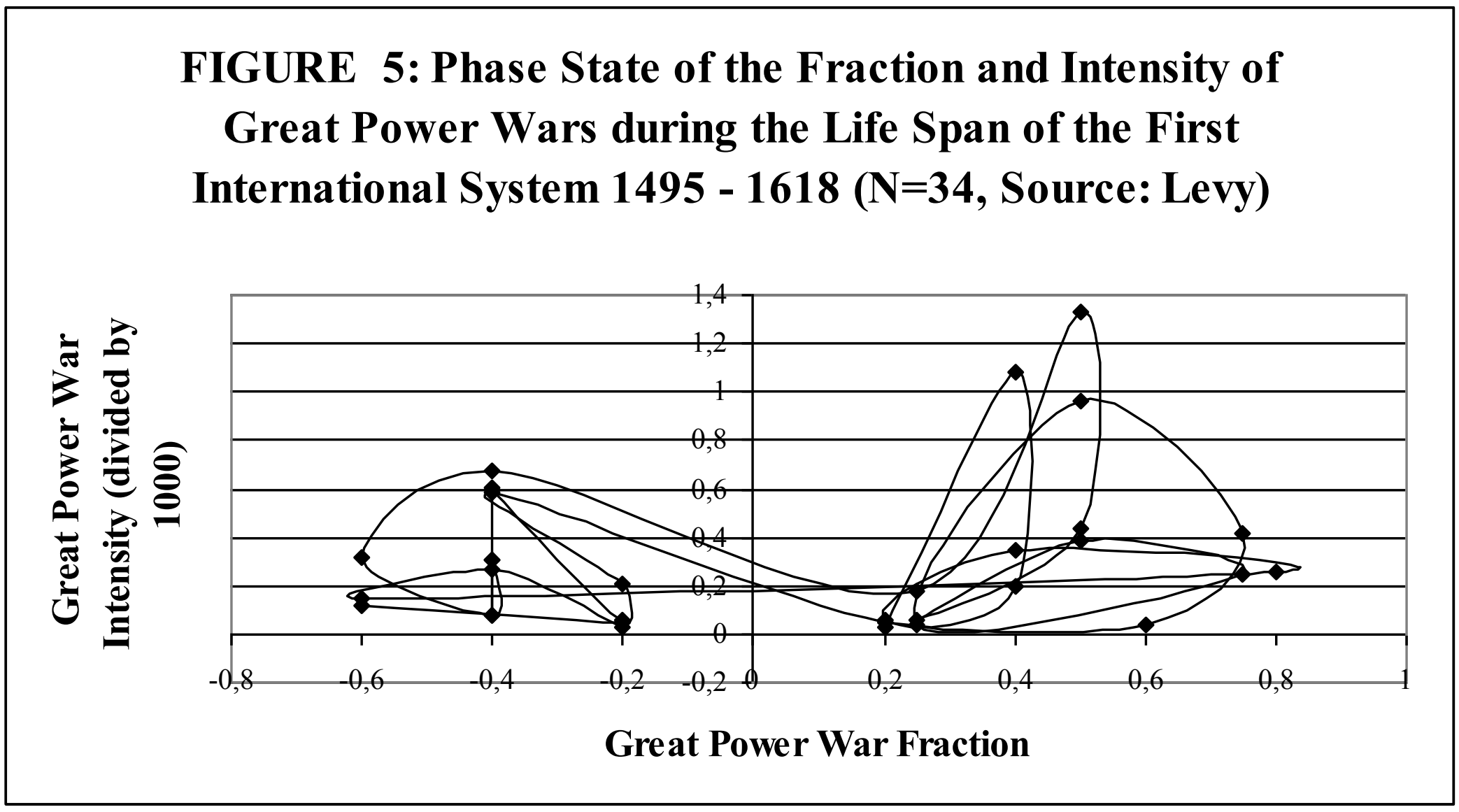

*Figure 1.5. This figure shows the circular trajectories in the phase state of consecutive Great Power wars during the life cycle of the first international system (Piepers, 2007)*

In next figure the zigzag pattern in the phase state during the exceptional period (1657 - 1763) is clearly visible. During the subsequent time span, circular trajectories again appear in the phase state of the International System.

[10] In order to visualize the directions of these trajectories we have given the fractions of wars constituting clockwise trajectories a negative value; these trajectories are shown in the second (left) quadrant of this figure.

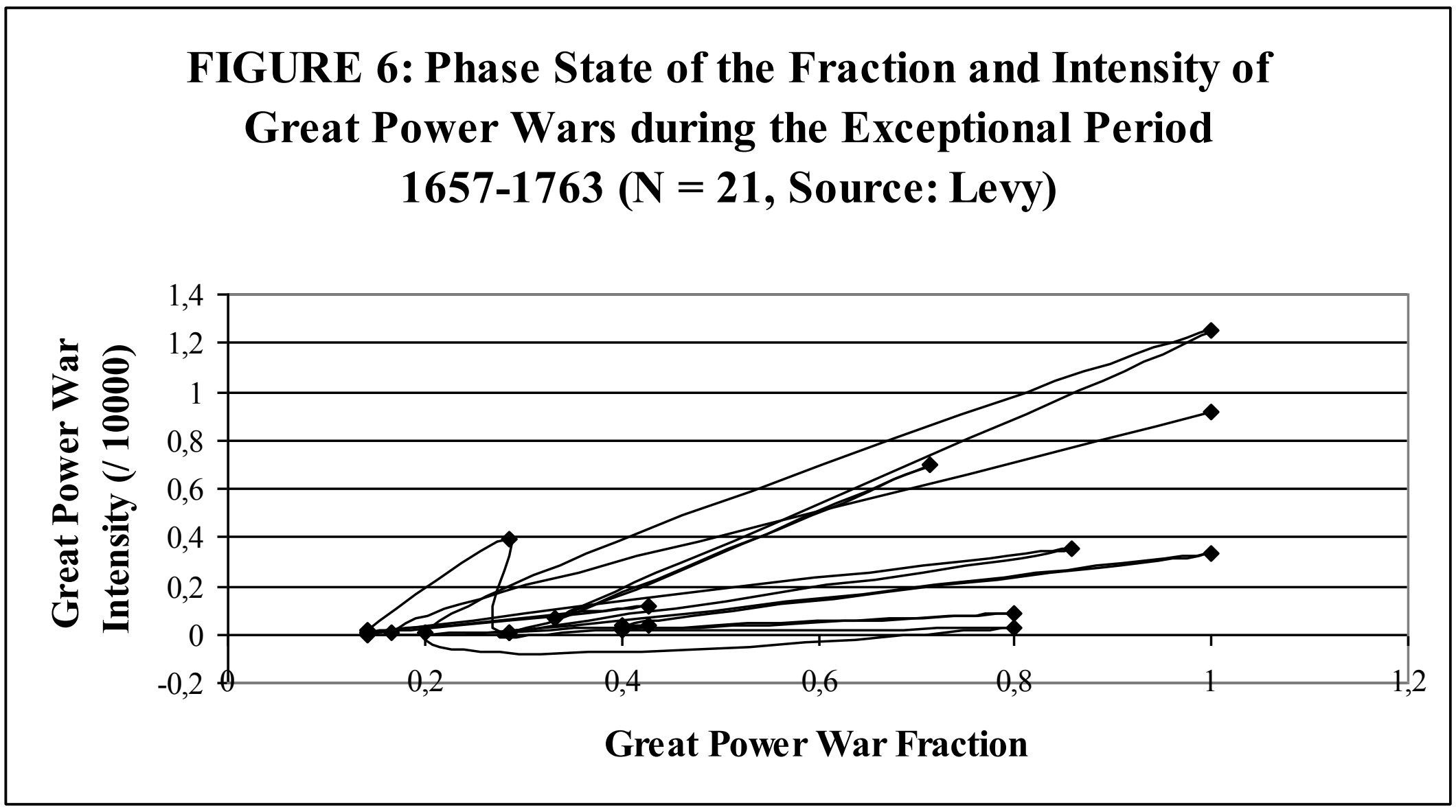

*Figure 1.6. Distorted war dynamics during the exceptional period (1657 -1763) (Source: Piepers, 2007)*

We assume - this is a hypothesis - that the clockwise and counter-clockwise trajectories indicate the existence of a chaotic attractor, influencing the war dynamics of the International System. Chaotic systems are deterministic systems, with a great sensitivity for the initial conditions of the system, and with highly unpredictable dynamics as a result (Lorenz, 1993, pp. 111-160, Strogatz 1994, pp. 320). The unpredictable dynamics of these systems are a result of the interplay between at least three degrees of freedom (Strogatz, 1994, pp.10).

Deviations from these trajectories (here, we are not referring to the zigzag patterns during the exceptional period, these are the consequence of a fundamentally other dynamic, as we will explain later) can - is our assumption - be contributed to the influence of stochastic events and perturbations: A chaotic attractor has a significant impact, but is not the only 'factor' influencing the war dynamics on the shorter term.

It is remarkable that the average fractions and intensities of Great Power wars constituting the respective trajectories develop very regularly, according to a certain logic (Piepers, 2007).

An interesting question is, if the zigzag dynamics during the exceptional period can be explained from a historical perspective as well. This seems to be the case.

During the period 1657 - 1763, the interactions and dynamics of this international system - historians agree - were to a high degree dominated by the intense rivalry between Britain and France: Both Great Powers were maneuvering and fighting for a hegemonic position in Europe. In 1763, the end of the Seven Years War, Britain finally achieved supremacy (Schroeder 1994, pp. 3-11). The war dynamics of the International System possessed a certain predictability during the exceptional period. As a consequence of this intense rivalry - it can be argued - international relations were 'simplified' and to a high(er) degree 'organized'.

During this timeframe decisions of other Great Powers to start or to participate in Great Power wars only depended on the position and actions of two actors: Britain and France. The positions and actions of other Great Powers were of minor importance and not taken into consideration, it seems.

From a system perspective - it can be argued - this intense rivalry resulted in a decrease of the number of degrees of freedom of the system to two. As a consequence of the intense rivalry between Britain and France, the chaotic attractor was replaced by a quasi-periodic attractor, temporarily dominating the war dynamics of the International System (Piepers, 2007). Quasi-periodic attractors cause more regular and predictable dynamics.

We argue that this temporary distortion disrupted the 'normal' development - life cycle - of the second international system as well, resulting in a lengthening of its life span (Piepers, 2006b).

Further research is necessary; the findings discussed in this paragraph are only the result of a superficial investigation into these remarkable dynamics.

## 6. Stability, resilience and social consolidation.

In this paragraph we discuss the concepts of stability and resilience of the international system, and show how wars seem instrumental in a process of social consolidation, at least during the time frame 1495-1945.
As explained, each punctuation - systemic war - during this time frame resulted in a qualitatively different international system, with its typical arrangements - rules and institutions - regulating and influencing the (inter)actions of states constituting the (new) system.
At a closer look it becomes clear that the four consecutive systems have not only specific qualitative, but also some remarkable quantitative characteristics.
We have defined two important properties of international systems: Their stability and resilience. This approach is based on ecosystem research (Pimm, 1991, Gunderson, 2002, pp. 25-62).
Both properties are - as we will show - useful viewpoints to get a better understanding of the dynamics and development of the International System on the longer term.
We have defined *stability* as the ability of an international system to sustain itself in a condition of relative rest. The war frequency and status dynamics of consecutive international system are measures for the stability of these systems and indications for its development on the longer term: See table 2.

| **Stability of international systems** | | |
|---|---|---|
| International system | War frequency | Status dynamics |
| 1 | 0.37 | 8 |
| 2 | 0.24 | 5 |
| 3 | 0.17 | 3 |
| 4 | 0.05 | 0 |

*Table 1.5. Quantification of the stability of consecutive international systems by measuring their respective war frequencies and level of status dynamics.*

The *war frequency* of international systems is calculated by dividing the number of Great Power wars during a life cycle of an international system by its lifespan. The lifespan of an international system is calculated by determining the difference between the start year of the punctuation that ends the life cycle of that particular system, and the end year of the preceding punctuation. The calculations of these system variables are based on Levy's dataset (Levy, 1983)[11].
These calculations show that the war frequencies of consecutive international systems decrease nearly linearly: See below figure. This implies - in accordance with the definition of stability in this context - a linear increase in the stability of consecutive international systems.

[11] Great Power wars outside the European continent with only one European participant are excluded from this overview. It concerns eight wars in the period from 1856 until 1939. This is a fundamentally different category of wars, indicative for the globalization of the International System, and obscuring the process of social expansion in Europe. Great Power wars constituting punctuations are excluded as well, because of their different function.

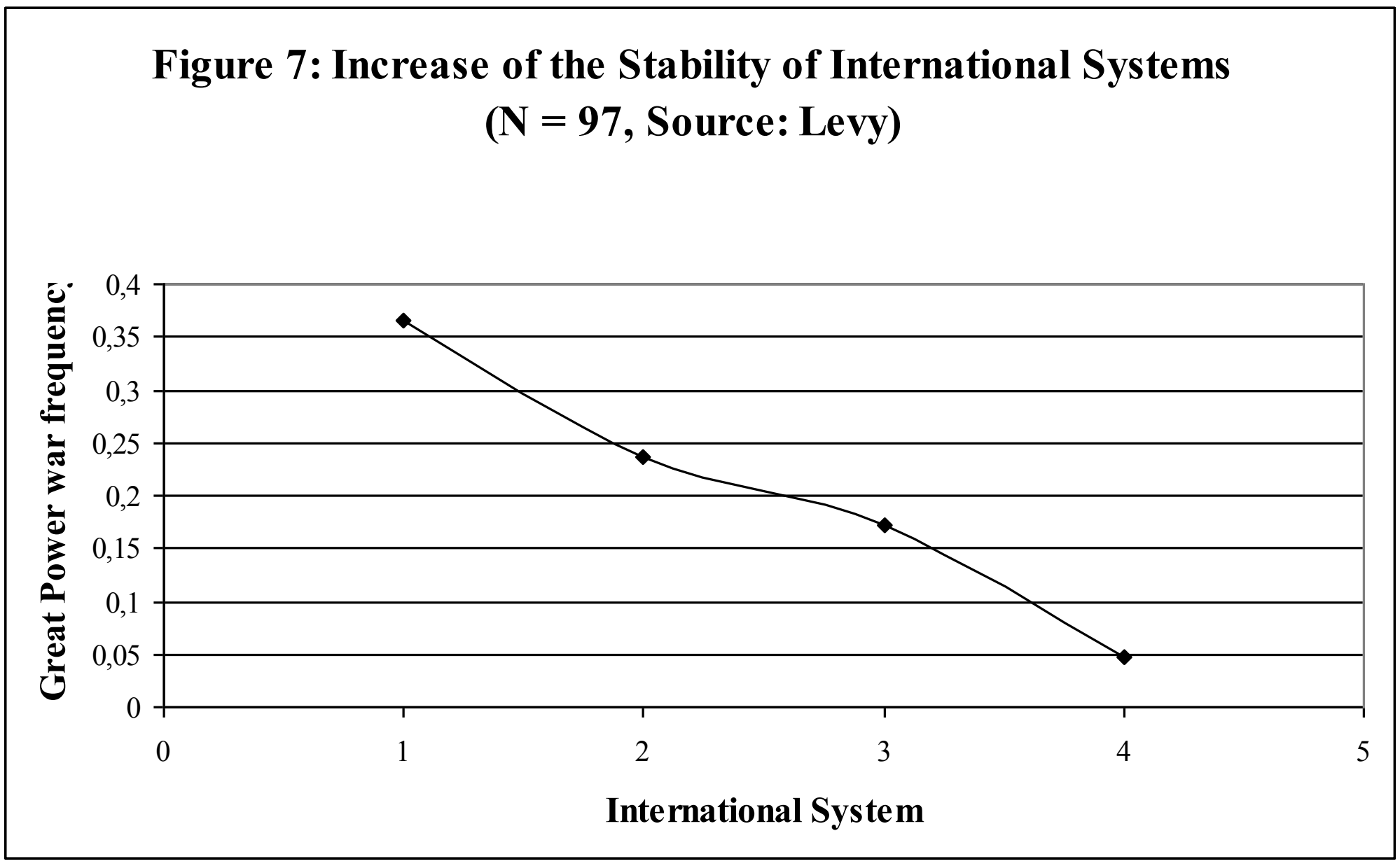

*Figure 1.7. Linear increase of the stability of consecutive international systems (Source: Piepers, 2007)*

The, what we call, 'status dynamics' of the International System are another indication of the development of the stability of the International System. The *status dynamics* of the International System concern the number of states that acquire or lose their Great Power status.

During the first four international systems, respectively eight, five, three, and zero status changes occurred; status changes during systemic wars are excluded (Levy 1983, pp. 47). It is interesting to note - and possibly no coincidence - that most status changes occur during the life span of international systems and not during systemic wars (Piepers, 2007). This finding supports the assumption that Great Power dynamics contribute to the build up of inconsistencies in the system, and systemic wars to correct these inconsistencies.

Two of the three status changes during the third international system concerned the United States (1898) and Japan (1905). These findings not only emphasize the increase in stability of the (European) International System, but also make us aware of the increased impact of non-European states on the dynamics of the International System.

We have defined *resilience* of an international system as the ability of an international system to sustain itself within a particular stability domain, an international system in the context of this research (Gunderson, 2002, Piepers, 2007).

The number of Great Power wars that is required to 'push' an international system out of its stability domain ('into' systemic failure) and the life span of international systems are indications for the resilience of international systems. See table 3.

| **Resilience of International Systems** | | |
|---|---|---|
| International system | Number of Great Power wars | Life span (years) |
| 1 | 45 | 123 |
| 2 | 34 | 144 |
| 3 | 17 | 99 |
| 4 | 1 | 21 |

*Table 1.6. Quantification of the resilience of consecutive international systems by measuring the number of wars required getting to the 'next' systemic war, and by measuring the life span of these systems.*

The decrease of the number of Great Power wars required to push consecutive international systems out of their respective stability domains shows that the *resilience* of the international system has decreased over time. Again, an almost linear relationship can be identified: See figure 8.

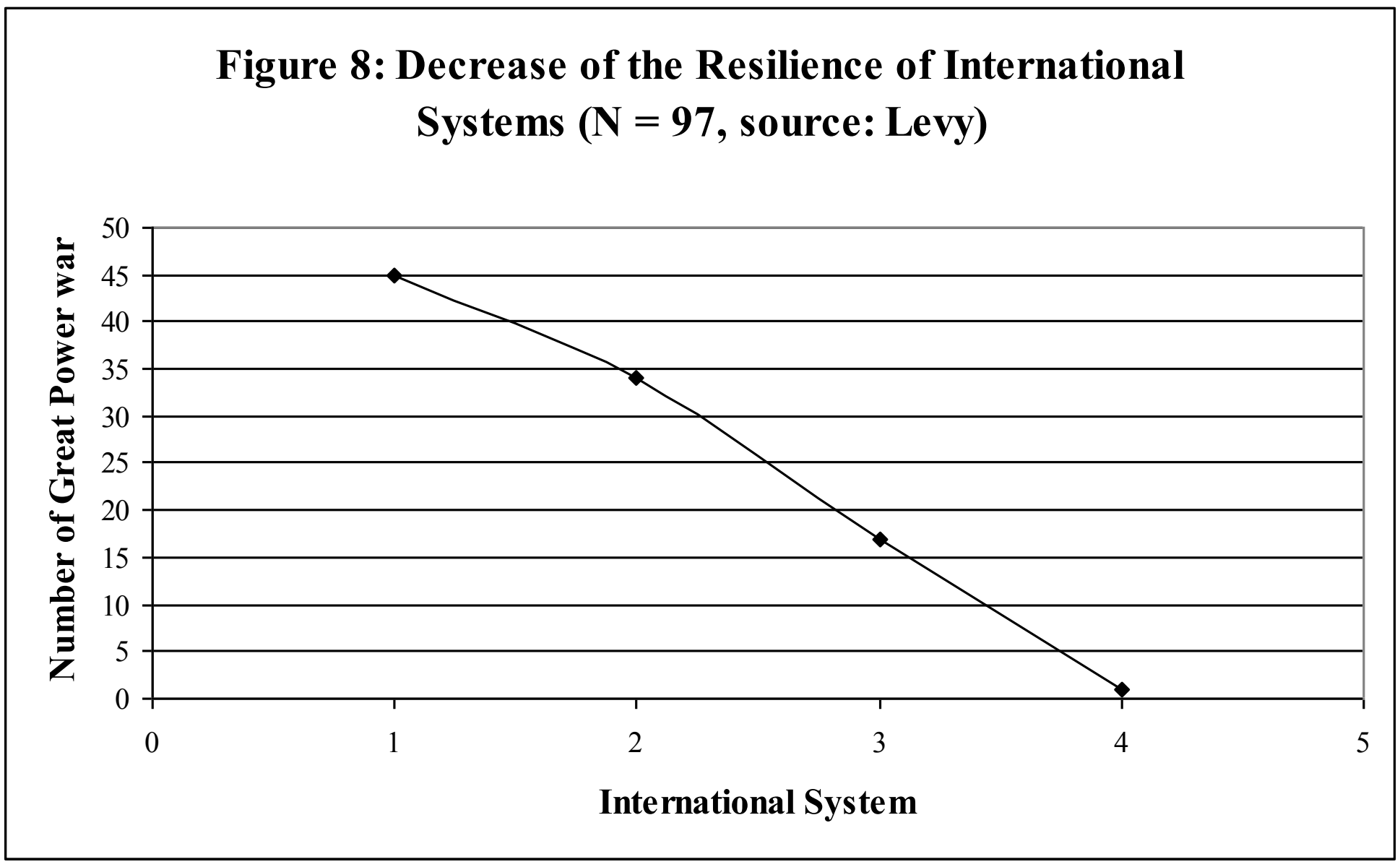

*Figure 1.8. Linear decrease of the resilience of consecutive international systems (Source: Piepers, 2007)*

Another indication for the development of the resilience of international systems is the decrease in the lifespan of these systems. The figures show - except for the second international system - a decrease of the lifespan of consecutive systems[12].

We argue that, the (linear) increase in the stability of the International System, as a consequence of the war dynamics of the system, can be understood as a process of 'social consolidation' (Piepers, 2008).

## 7. A consistent framework.

From our research, a consistent framework emerges: The self-organized critical (SOC-) characteristics, the punctuated equilibrium (PE-) dynamics, chaotic war dynamics, and the 'steady' development of the International System towards a condition of increased stability - seem to be closely related. SOC results in a PE-dynamic. Systemic wars enable the release of 'system-sized' tension and frustration, that were able to build up as a consequence of a network effect, halfway the life span of an international system. Systemic wars constitute fundamental reorganizations of the International System. SOC and PE-dynamics seemed to have been functional in a process of social consolidation in Europe.

It is interesting to note, that in the International System order and chaos seem to go hand in hand, however at different levels of analysis. SOC-characteristics and the PE- dynamics of the International System result in a certain order and predictability in the development of the International System on the longer term, while at the level of consecutive Great Power wars, a chaotic attractor seems to influence the war dynamics, resulting in a high degree of unpredictability.
We assume that the typical chaotic dynamics were disrupted two times since 1495: During the period 1657 - 1763, as a consequence of intense rivalry between Britain and France, and during the period from 1945 - 1991, as a consequence of intense rivalry between the United States and the Soviet Union. These distorted war dynamics - we argue - resulted in a certain ossification of these particular international systems, disturbing their 'normal' life cycle, and lengthening their life-span.

[12] We assume that the relatively long lifespan of the second system is related to the erratic- non chaotic - war dynamics during the exceptional period 1657 - 1763. It seems that the simplified dynamics during this period have resulted in a lengthening of the lifespan of this system.

**APPENDIX: Watts' model**

In *a simple model of global cascades on random networks* Watts' researches the origin of "large but rare cascades that are triggered by small initial shocks". This research and its findings possibly provide some clues for a better understanding of the war dynamics of the International System.
Watts identifies two regimes in which the network in his simulations is susceptible to very large (so called 'global') cascades: a regime in which the cascade size is limited by the connectivity of the network, resulting in a power law distribution of cascade sizes, and a regime when the network is highly connected, cascade propagation is instead limited by the local stability of the nodes themselves. In case of the second regime the size distribution of cascades is bimodal.
In the first regime – Watts notes – it is found that the most connected nodes are far more likely than average nodes to trigger cascades, but not in the second regime (Watts, 2002, 5766).
We do not suggest that the International System works exactly as this "*simple model of global cascades on random networks*". However our research suggests that probably some similarities in behavior exists. Validation is required, with special focus on the isomorphism of these systems.
In his approach to this phenomenon Watts uses the concept of vulnerable clusters (Watts, 2002, 5767). According to this approach, states (comparable to vertices (nodes) in Watts his research) can have a vulnerability condition as well, and can be stable or unstable. In the last case - for example – in a one-step or two-step sense, depending on the number connections required to change position (e.g. from cooperation to conflict), to change the position of the state from cooperation to conflict.

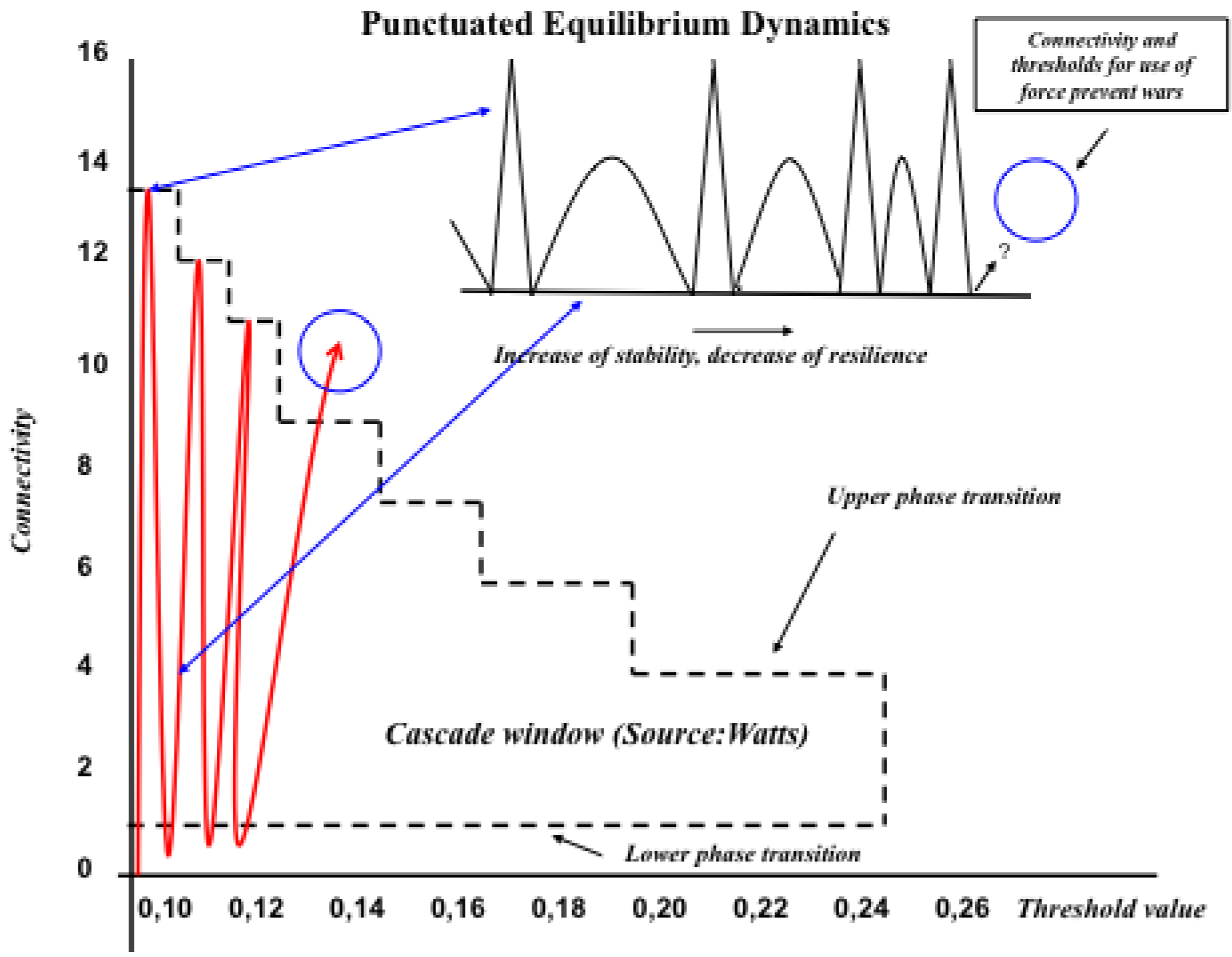

*Figure 2.1. Punctuated Equilibrium Dynamics in relation to Watts' framework (Source: Piepers, 2007)*